# Code smells: A Synthetic Narrative Review

Peter Kokol Prof
*University of Maribor, Slovenia*, peter.kokol@um.si

Marko Kokol
*Semantika*, marko.kokol@um.si

Sašo Zagoranski
*Semantika*, saso.zagoranski@semantika.si

# Code smells: A Synthetic Narrative Review

Peter Kokol [1*], Marko Kokol [2], Sašo Zagoranski [3]

[1] University of Maribor, Faculty of Electrical Engineering and Computer Science, Žitna Ulica, 2000 Maribor, Slovenia
[2] Semantika, Zagrebška cesta 40a, 2000 Maribor, Slovenia
[*]peter.kokol@um.si

**Abstract: Code smells are symptoms of poor design and implementation choices, which might hinder comprehension, increase code complexity and fault-proneness and decrease maintainability of software systems. The aim of our study was to perform a triangulation of bibliometric and thematic analysis on code smell literature production. The search was performed on Scopus (Elsevier, Netherlands) database using the search string "code smells" which resulted in 442 publications. The Go-to statement was the first bad code smells identified in software engineering history in 1968. The literature production trend has been positive. The most productive countries were the United States, Italy and Brazil. Eight research themes were identified: Managing software maintenance, Smell detection-based software refactoring, Architectural smells, Improving software quality with multi-objective approaches, Technical debt and its instance, Quality improvement/assurance with mining software repositories, Programming education, Integrating the concepts of anti-pattern, design defects and design smells. Some research gaps also emerged, namely, New uncatalogued smell identification; Smell propagation from architectural, design, code to test, and other possible smells; and Identification of good smells. The results of our study can help code smell researchers and practitioners understand the broader aspects of code smells research and its translation to practice.**

## 1. Introduction

Many modern Computer Science approaches draw their inspiration from nature [1]. Smells play an important role in communication and assessments of other beings and objects, mainly in mating rituals and searching for food. Fowler [2] defined code smells as symptoms of poor design and implementation choices. Such symptoms may originate from activities performed by developers during emergencies, poor design or coding solutions, by making bad decisions, or employing so called anti-patterns [3]. Code smells could also be the consequence of so-called technical debt [4]. Among other things, they might hinder comprehension [5] and increase code complexity and fault-proneness and decrease maintainability [6]. Another viewpoint is that code smells are less than optimal decisions during software development and implementation, resulting in lower quality, low maintainability, hard to evolve and increased maintenance costs [7–9]. To overcome the above problems, code smells must be identified and dealt with [10]. Identification of code smells relies on structural information extracted from the source code [11]. However, detecting code smells is a complex activity, which depends on human factors [12], programming language [13] and similar.

Sharma and Spinellis [14] surveyed the research on software smells using the conventional review approach and Kokol et al performed a brief synthetic scoping review [15]. In our study, we used another approach, namely bibliometric analysis and bibliometric mapping triangulated with thematic analysis. In that manner, we induced a so- called synthetic narrative review [16, 17]. Contrary to ordinary reviews which result in Tables of evidence, bibliometric mapping visualises the content of research publications in the form of various scientific landscapes [18]. These landscapes can be analysed further using different interpretative approaches which than reveals a different perspective on a research area. Our objective was to analyse trends of the code smells` research literature production, its historical roots, its distribution and, finally, to identify main research themes and directions.

## 2. Methods

Bibliometrics [19] has its origins in the beginning of the last century. However, it became »operational« in 1964 with the introduction of the science citation index, and prominent because of the need to measure the effects of the large investments going into the research and development [20, 21]. Bibliometrics analyses the properties of literature production in terms of measures, like the number of articles in a scientific discipline, trends of literature production, most prolific or productive entities, most cited papers and authors, etc.

An interesting technique used in bibliometric analysis is bibliometric mapping, which visualises literature production based on various text mining techniques [18]. A popular bibliometric mapping software tool is the VOSviewer (Leiden University, Netherlands) [22]. VOSviewer software extracts, analyses and maps/visualises terms, keywords, authors, countries and other bibliometric entities in the form of different science landscapes . Another recent tool often used in bibliometric analysis is CitedReferenceExplorer used to perform Reference Publication Year Spectroscopy (RPYS). It is an open source software used to analyse, tabulate and visualise the cited references found in a corpus of research publications (www.crexplorer.net). Its primary aim is to identify those publications which have been referenced most frequently, and is, thus, suitable to identify historical roots of specific research fields [23].

### 2.1. Data source and corpus

The search was performed on Scopus (Elsevier, Netherlands), the largest abstract and citation database of peer-reviewed literature: Scientific journals, books and conference proceedings. The corpus was formed on September 24th, 2018, using the search sting "code smells" in information source titles, abstracts, and keywords on all publications covered by Scopus.

### 2.2. Data extraction and analysis

Using Scopus analysis services, we exported authors` affiliation details, source title, publication type, abstracts, titles and publishing years to MS Excel (Microsoft, USA), VOSviewer and CitedReferenceExplorer, where they were analysed. All common terms, like study, baseline, control group, trend, method, significance, country and city names, time stamps were excluded from the analysis. Synonyms were mapped into one entity, for example, bad smell(s), bad code smell(s), code smell(s) into code smells, refactoring and software refactoring into software refactoring, and machine learning and machine learning techniques into machine learning. Three maps were induced (1) Country co-author network – timeline coloured, based on the average publishing year (2) The authors` keyword co – occurrence clustered science landscape and (3) Term clustered science landscapes for the period 2017-2018. The keywords` landscape was analysed by thematic analysis [24] to determine the main research themes. The term science landscape was used to reveal current research directions and possible gaps in research. The cited references were analysed using CitedReferenceExplorer to identify the historical roots of the code smells` research.

## 3. Code smell research

The search resulted in 442 publications. Among them there were 309 conference papers, 87 articles, 4 reviews, 41 conference reviews and 1 book chapter. The first two publications indexed in Scopus were published in 2002 at the Conference on Reverse Engineering. One was the proceedings` introduction [25] and the other the proceeding paper about detecting code smells during inspections of code written in Java [26]. The first slight rise in research literature productivity was noticed in 2005, and the next larger one in 2009. The exponential growth of the production started in 2014 The exponential growth of the production started in 2014 reaching its peak value in 2017 with 98 publications.

### 3.1. Historical roots of code smells research

RPYS analysis (Fig- 1.) showed that the code smells` historical roots could be categorised in four periods:

- First bad code smell identified (1968): In 1968, Edgar Dijkstra published his famous letter [27], declaring Go-to statements as harmful and, in that way, he probably identified the first bad code smell in software engineering history;
- Software Metrics era (1969-1982): The next era was associated with the beginnings of Software Metrics` development [28–32]. Allen [28], analysed the control flow relationships, and expressed them in a directed graph in a manner to optimise coding. The theory of Software Science [29] applied the methods of science to the complex problem of software production, and validated it with experimental evidence. Despite its controversy, it received widespread attention from the Computer Science community and initiated the development of Software Metrics. The idea of Software Metrics was then implemented by a program called checker Lint, which examined C source programs for bugs and violation of the type rules [30] and, later, with the experiment investigating how different types of modularization and comments are related to programmers' ability to understand programs [31];
- Program restructuring/refactoring era (1983-1993): This era begun with the Guimaraes paper [33], in which he presented recommendations on how to reduce program maintenance expenditures based on the analysis of application portfolios and personal interviews with top computer executives and systems development personnel. The research continued with the development of laws of software evolution [34], and definition, development and automation of software restructuring and refactoring related to improving maintainability [35–37];
- Applying metrics to the maintainability era (1994-2007): This era started with a book presenting specific object-oriented metrics derived from several actual projects. Metrics has been applied in real world situations [38]. It was followed by a paper presenting a minimal set of easily calculated metrics which, as a whole, supported maintenance quantitatively and, consequently, decision-making in the program development. [39, 40]. Another important milestone occurring in this period was the introduction of patterns and anti-patterns in 1995 by Koenig [41]. Koenig defined anti-patterns as "An anti-pattern is just like a pattern, except that, instead of a solution, it gives something that looks superficially like a solution but isn't one".
- Code smells era (2008-2018). This last era was concerned with research directly related to code smells, their detection, and their relation to fault proneness [42–45].

### 3.2. Distribution of the code smell research

Publications appeared in 61 different source titles. The most prolific source titles are listed in Table 1. Most of the top prolific source titles are proceedings of internationally established conferences or recognised journals.

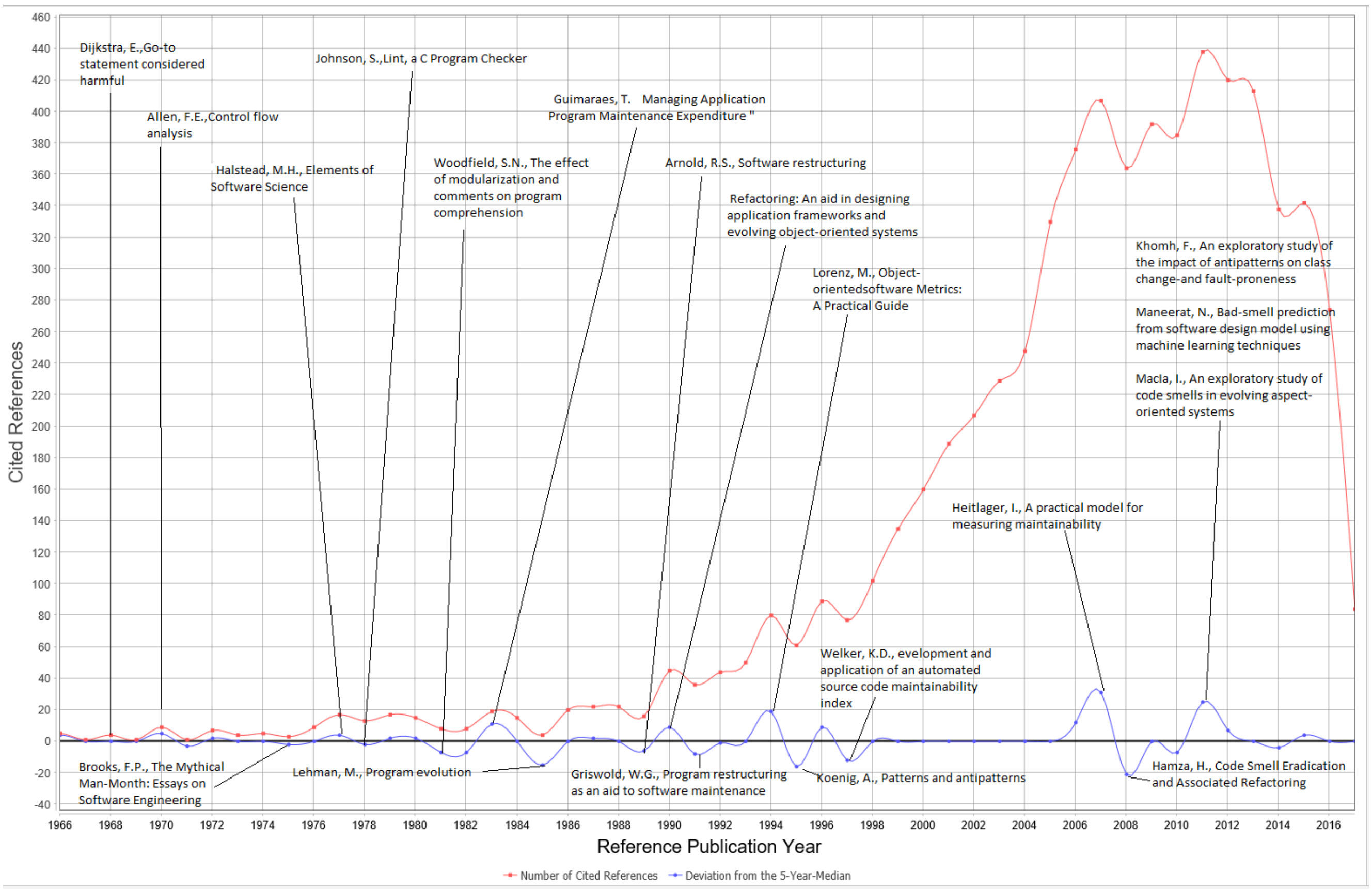

***Fig. 1.**. RPYS analysis of the publications cited in the Code smells` research*

**Table 1** The 10 most prolific source titles

| Source title | Number of publications |
|---|---|
| Proceedings International Conference on Software Engineering | 30 |
| Lecture Notes in Computer Science Including Subseries in Artificial Intelligence and Bioinformatics | 26 |
| ACM International Conference Proceeding Series | 18 |
| IEEE International Conference on Software Maintenance | 13 |
| IEEE Transactions on Software Engineering | 13 |
| Journal of Systems and Software | 10 |
| Information and Software Technology | 9 |
| Empirical Software Engineering | 8 |
| CEUR Workshop Proceedings | 7 |

The research was distributed over 50 countries and 172 institutions. The 10 most productive countries are shown in Table 2. The United States was the most productive country, followed by Italy and Brazil. The top 10 countries produced almost 83% of all research literature regarding code smells. Among them there were five G7 countries, missing are Japan and the United Kingdom, which are ranked 12th and 13th. With the exception of Africa (South Africa, Nigeria and Tunisia produced one paper and were ranked 37th) and Australasia (Australia is 30th), all other continents are represented in the list of top productive countries. Hence, the research on code smells seems to be globally widespread, despite the fact that most of the research is performed in the most developed countries.

**Table 2** The 10 most productive countries

| Country | Number of publications |
|---|---|
| United States | 93 |
| Italy | 53 |
| Brazil | 49 |
| Canada | 37 |
| Germany | 35 |
| India | 26 |
| Netherlands | 24 |
| Norway | 17 |
| Switzerland | 16 |
| China | 16 |

Authors from 34 countries cooperated on a co-authorship basis (Figure 2.). The USA was the most productive country regarding co-authorships (n=58), followed by Italy (n=45), Canada (n=30), Germany (n=20) and Switzerland (n=19). The strongest co-operation was between the USA and Italy (n=14). Italy and Switzerland (n=8) and the USA and Canada (n=6). Based on the average year of publications Germany and the UK seem to be the countries to start the research on code smells (violet colour), followed by the USA, Canada, Norway and Poland (blue colour). Late starters were China, Japan, Ireland and Spain (green and yellow colours).

The most productive institutions are presented in Table 3. They represent around. 43% of total research production. While the most productive country is the USA, it is interesting to note that, among the top 13 productive institutions, none is from the USA, and most (n=5) are from Italy and Europe in general (n=9). That might indicate that the research in code smell is widespread in the USA but is less centred, contrary to other continents, where the research seems to be more concentrated in strong centres.

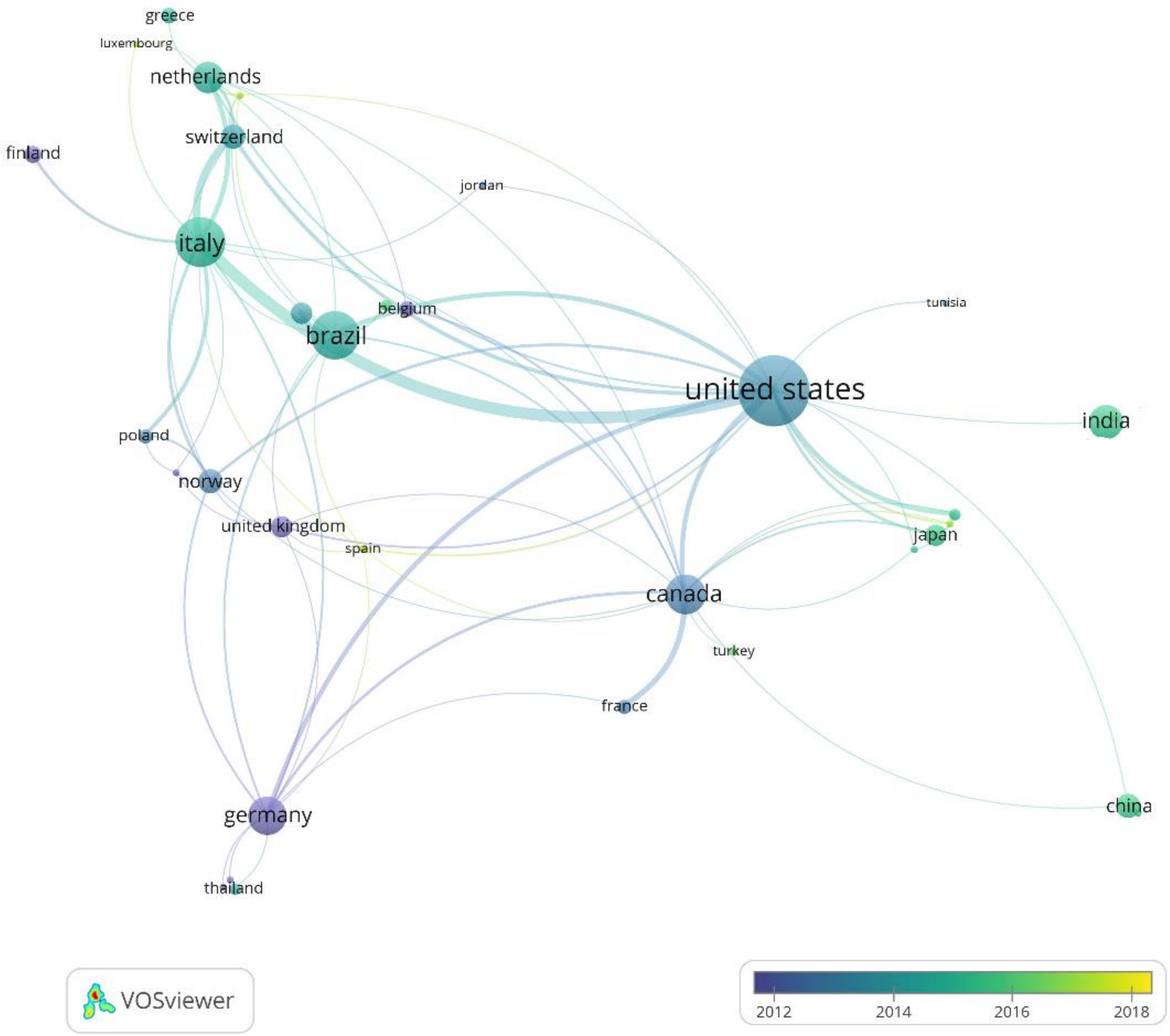

***Fig. 2.*** *Country co-operation network based on co-authorship*

**Table 3** Most productive Institutions

| Affilation | Number of publications |
|---|---|
| Pontificia Universidade Catolica do Rio de Janeiro | 25 |
| Universita degli Studi di Milano - Bicocca | 23 |
| Delft University of Technology | 21 |
| Universita di Salerno | 19 |
| Universita degli Studi del Molise | 16 |
| Universita degli Studi del Sannio | 12 |
| Universite de Montreal | 12 |
| Universita della Svizzera italiana | 11 |
| Ecole Polytechnique de Montreal | 11 |
| Universidade Federal da Bahia | 10 |
| Brunel University London | 10 |
| Simula Research Laboratory | 9 |
| Free University of Bozen-Bolzano | 9 |

### 3.3. Code smells` research themes

The author keyword network is presented in Figure 3. Based on cluster terms, we named and defined code smells` research themes using thematic analysis, and located and narrated representative publications for each theme. Eight research themes were identified after integrating terms from yellow and brown clusters into one theme:

- *Manging software maintenance* (orange colour), is concerned with improving maintainability of code by employing software metrics, code reviews and source code analysis. Code smells are assumed to indicate bad design that leads to less maintainable code. However, a controlled experiment including six professional software developers showed that the effects of the 12 code smells on maintenance effort were limited [46]. On the other hand, a study of 28 novice developers showed that collaborative efforts in code review,s like Pair Programming, improved the identification of code smells and improved maintainability [47]. Additionally, Zazworka [48] showed that code smells correlate with technical debt in a different way than grime build up or modularity violations, and can be used to assess change proneness and, thus, maintainability.
- *Smell detection-based software refactoring* (red colour) research deals with the use of various code smell detection techniques and strategies, like machine learning and software visualisation, to optimise software by removing bad smells employing software refactoring. Refactoring is a key issue to increase internal software quality and maintainability. Code smells are used to identify structures where refactoring should be applied [49]. Removing duplicate codes or code clones in Erlang programs was one of first attempts of automatic refactoring [50]. Sometimes, multi-objective models which maximise the trade-off between quality improvements, severity and importance of refactoring opportunities, should be used for refactoring [51]. In some instances, tools can be used to support automatic refactoring. JMove, a publicly available tool comparing the similarity of the dependencies established by a method with the dependencies established by the methods in possible target classes, is one of them [52]. JSpIRIT is a tool which prioritises code smells based on customizable detection strategies based on the configuration of their manifold criteria [53]. FaultBooster identifies problematic code parts using static code analysis, and running automatic algorithms to fix selected code smells [54]. Liu et all [55]developed a monitor-based refactoring framework which analyses changes in the source code instantly and warns programmers if the changes have the potential to introduce code smells, and should be refactored. Instead of automatic refactoring, code smells` detection could be used just for identification of refactoring opportunities. Imazato [56] used machine learning on development histories to identify code to refactor, taking into account specific project attributes. Code smell severity is an important factor when prioritising refactoring efforts. Thus, Arcelli and Zanoni [57] used different machine learning models to determine the severity of code smells. Software visualization is another technique to identity refactoring opportunities. Interactive ambient visualization was employed to help programmers identify and interpret code smells in the so-called soft advice process supporting software development [58]. It was also used to detect smelly formulas in spreadsheets which were used in business applications [59]. Another application of identification code smells trough software visualization was used to detect Long Method, Large Class and Long parameter list smells [60]. Combining visualization with crowd-smelling employs a concept collaboration of a global community of programmers which contribute smell detection algorithms, which are then visualised in smelly maps to help programmers in the manner to improve detection accuracy and mitigation of specific problems [61]. Despite all the research presented above, a large empirical study performed on change histories of 200 open source projects revealed that most of the code smells were introduced into the code by its creation, and not during later development phases. Despite software evolution, four fifths of smells remain in the code. Among the rest of smells, only one tenth is removed by refactoring [62].
- *Architectural smells`* (Light blue colour) research combines the principle of Aspect and Object-oriented Programming (AOP) with the detection of architectural smells during software architecture design. Architectural smells are different to anti-patterns or code smells, and are defined as "frequently recurring software designs that can have non-obvious and significant detrimental effects on system lifecycle properties« [63]. A developer using AOP sometimes unwittingly inserts architectural smells into architectures which can later cause modularity problems, and some of these smells are even not targeted by refactoring detection strategies [64]. Architectural smells might be more critical than code smells and even harder to detect and refactor. Arcan is a tool developed to overcome these problems. It detects architectural smells through the analysis of architecture dependency issues [65].
- *Improving software quality with multi-objective approaches* (rose colour) is involved with balancing different trade-offs using search-based software engineering, introducing design patterns and analysing inter-smell relations. MORE is an automatic refactoring tool balancing the trade-off`s three objectives, namely: Improving design quality, fixing code smell and introducing design patterns. Search- based software engineering, more precisely the nondominated sorting genetic algorithm, was adopted to solve this problem [66]. In another study, multiple-objective genetic programming was used to determine the best combination of metrics that maximise the detection of code smells and minimise the detection of well-designed code [67]. Inter-smell relations can cause maintenance problems or intensify the negative effects of single code smells, therefore it is necessary to identify coupled or collocated smells [46, 68].

- *The Technical debt and its instances (light rose colour)* theme is concerned with the research of various instances of technical debts (design debt, defect debt, etc.), and their impact on software quality. Technical debt is a metaphor associated with developer decisions regarding the trade-of between different dimensions of software development, for example, quality vs time [69], or the trade-off between long and short-term goals in software development [70]. While programmers are aware of technical debt, its concrete distances are hard to conceptualise and manage. It is shown that agile practices help to reduce debts [71]. Debt tracking is another strategy employed especially by large organisations [72], while game theory is used to reduce debts in cloud systems [73]. A frequent instance of technical debt is incomplete, temporary and buggy code that required rework and is called Self-Admitted Technical Debt (SADT) [74]. TEDIOUS (Technical Debt IdentificatiOn System) is a machine learning based approach which uses different structure, class and readability metrics to recommend developers when they have to "self-admit" technical debt [75]. SADT is also detected using natural language processing and text mining of source code comments [76, 77] It is interesting to note that new studies revealed that active God class, once believed to be amongst the most harmful code smells, can be differentiated as strong, stable and ameliorate [78].
- *Quality improvement/assurance with mining software repositories (violet colour)* in the context of code smells is focused on analysing large software archives empirically to find relations between code/test smells, dimensions of the coding process and attributes of code quality. Mining software repositories has been used, for example, to study the impact of code smells on the software change-process [43] or maintainability [79, 80], developing defect and bug prediction models [81, 82] and analyse code smell cooccurrences [7]. Mining software repositories have been also used in analysis of test smells, showing that there is a high diffusion of test smells in open and industrial software systems, and that test smells have a strong negative impact on comprehensibility and, consequently, maintainability [83, 84]. Similar to code smells, violations of testing principles are defined as test smells. The first two identified test smells were General Fixture and Eager Test [85],
- *Programming education* (blue colour) also has to deal with code smells. Block-based-programming languages, like Alice, Blocky and Scratch, have become increasingly popular in the education of novice programmers and also children. However, research showed that block-based programs are frequently smelly, [86, 87] and that software engineering issues like code smells, duplication and refactoring should be included in Scratch programming courses [88],
- *Integrating the concepts of anti-pattern, design defects and design smells* (green colour) research is associated with casual dependencies of the above three phenomena and how to detect them. Anti-patterns and design smells both have negative effects on program comprehensibility, maintenance and flexibility, and are the result of poor design [89, 90] Poor design results in design defects [91]. Design defect, design smell and anti-patterns have been detected with different techniques, like DÉCOR, a language to define the steps for the specification and detection of design smells [92], IDS, a system which considers object oriented design as a living creature and detects design smells using the artificial immune system approach [93], Bayesian Belief Networks to detect anti-patterns [93], and FaultBuster [54], described above. A more recent study states that design flaws in object oriented programs may corrupt code and propose an automated design flaw detection, using multi pattern matching and detection rules reuse [94].

### *3.4. New Research Directions*

The comparison of the term science landscapes for the period 2015 – 2016 with the period 2017-1018 using the method presented by Kokol et al [95] did reveal themes presented below (the term landscape for the period 2017 – 2018 is shown in Figure 4):

- Code smells in mobile applications are especially harmful due to the exponential rise of new users and new applications. However, traditional code smell detection and refactoring approaches cannot be used directly in the context of IoS or Android apps` development. Thus, new tools and smells` catalogues are researched [96]
- Nano patterns, which are method-level traceable constructs and are more error prone and smelly than other patterns according to recent research. Thus, more intensive research is needed in this direction [97];
- Change patterns describe two or more files which are frequently changed together, either during development or maintenance. Change patterns increase fault proneness significantly, and more research is needed regarding their detection strategies and impact [98].
-

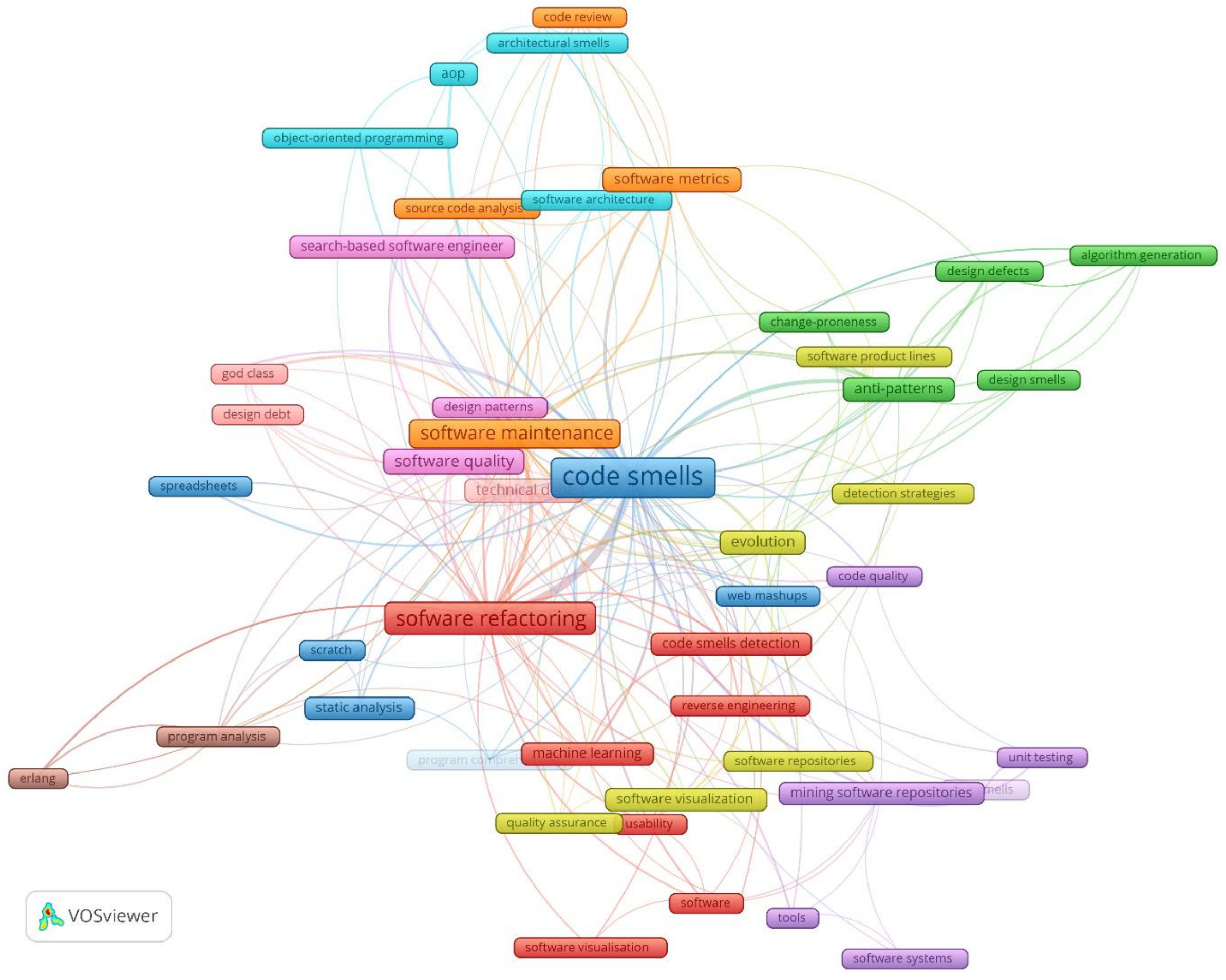

***Fig. 3.****Authors` keyword network*

## 4. Conclusion

Our synthetic narrative review presents a structural view of code smell research as revealed in research publications indexed in the Scopus database. The review can help code smell researchers and practitioners understand the broader aspects of code smells` research and its translation to practice. On the other hand, it can help a novice or a software engineer without specific knowledge on code smells to develop a perspective on the most important research themes and relations between them. Our study also revealed some research gaps which might present the challenges for future research:

- New uncatalogued smell identification;
- Smell propagation from architectural, design, code to test and other possible smells;
- Identification of good smells.

-

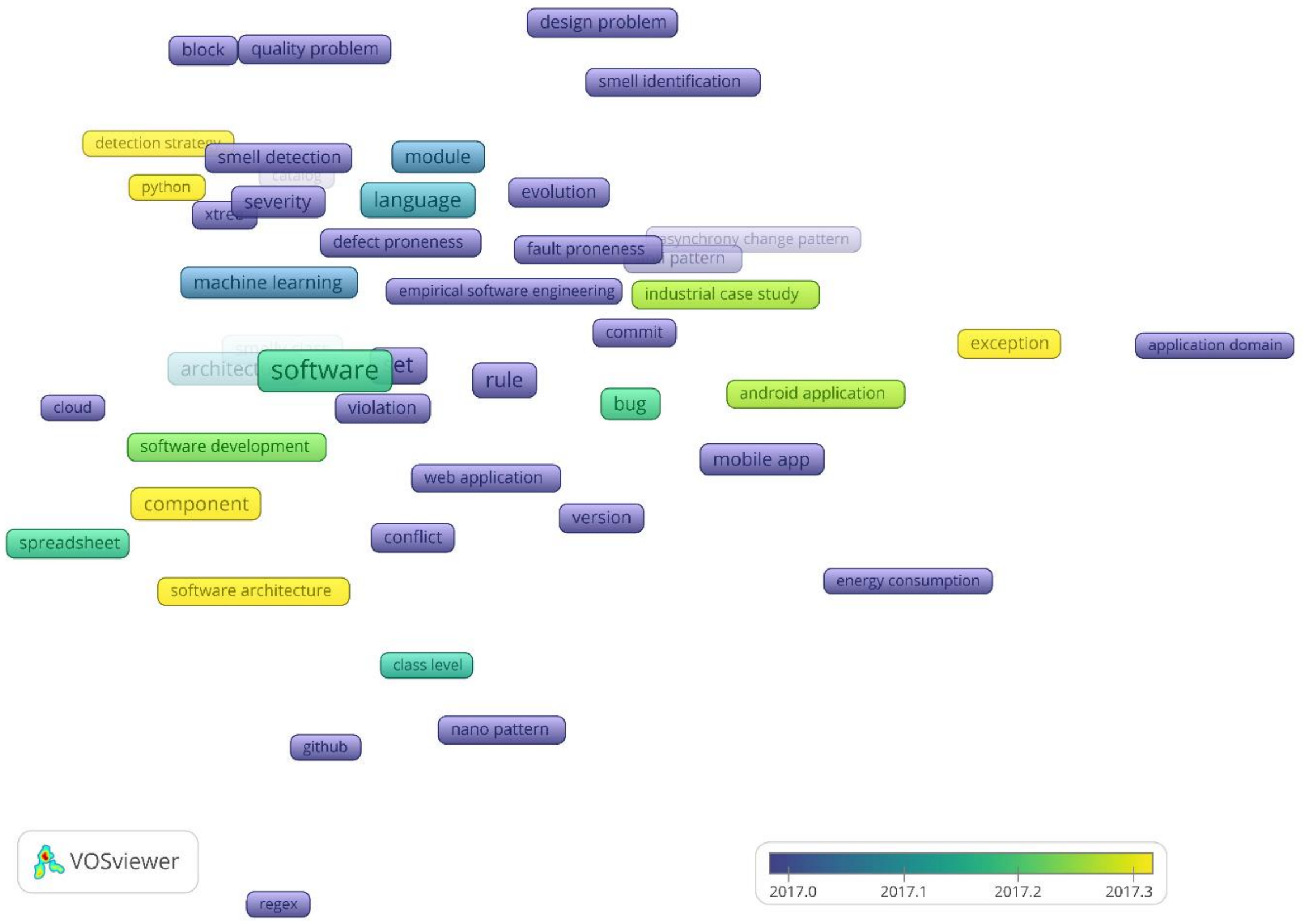

***Fig. 4.*** *The term "timeline landscape" for the period 2017-2018*